# Charge states, triple points and quadruple points in an InAs nanowire triple quantum dot revealed by an integrated charge sensor

Weijie Li,[1,2] Zhihai Liu,[3] Jingwei Mu,[1,2] Yi Luo,[1,4] Dong Pan,[5] Jianhua Zhao,[5] and H. Q. Xu[1,3,*]

[1] Beijing Key Laboratory of Quantum Devices, Key Laboratory for the Physics and Chemistry of Nanodevices, and School of Electronics, Peking University, Beijing 100871, China

[2] Academy for Advanced Interdisciplinary Studies, Peking University, Beijing 100871, China

[3] Beijing Academy of Quantum Information Sciences, Beijing 100193, China

[4] Institute of Condensed Matter and Material Physics, School of Physics, Peking University, Beijing 100871, China

[5] State Key Laboratory of Superlattices and Microstructures, Institute of Semiconductors, Chinese Academy of Sciences, P.O. Box912, Beijing 100083, China

[*] Author to whom correspondence should be addressed: hqxu@pku.edu.cn

ABSTRACT

A serial triple quantum dot (TQD) integrated with a quantum dot (QD) charge sensor is realized from an InAs nanowire via a fine finger-gate technique. The complex charge states and intriguing properties of the device are studied in the few-electron regime by direct transport measurements and by charge-sensor detection measurements. The measurements of the charge stability diagram for a capacitively coupled, parallel double-QD formed from a QD in the TQD and the sensor QD show a visible capacitance coupling between the TQD and the sensor QD, indicating a good sensitivity of the charge sensor. The charge stability diagrams of the TQD are measured by the charge sensor and the global features seen in the measured charge stability diagrams are well reproduced by the simultaneous measurements of the direct transport current through the TQD and by the simulation made based on an effective capacitance network model. The complex charge stability diagrams of the TQD are measured in detail with the integrated charge sensor in an energetically degenerate region, where all the three QDs are on or nearly on resonance, and the formations of quadruple points and of all possible eight charge states are observed. In addition, the operation of the TQD as a quantum cellular automata is demonstrated and discussed.

# 1. Introduction

Semiconductor quantum dots (QDs) made from narrow bandgap nanowires, such as InAs and InSb nanowires, are among the most potential platforms for the physical implementation of solid-state based quantum simulation and quantum information processing[1-6], due to the inherent properties of small effective masses, large Landé g-factors and strong spin-orbit interactions in the materials[7-13]. Coherent manipulations of spin qubit states via all-electrical means have now been achieved in InAs and InSb nanowire double QDs (DQDs) [14,15]. Experimental efforts have been made to extending DQDs to nanowire triple quantum dots (TQDs) [16-21], which are of great importance for achieving decoherence-free and exchange-only qubit operations[22-25] and coherent quantum teleportation[26,27], and for understanding rich physical phenomena, such as ground-state charge configurations at quadruple points (QPs) [16-19], quantum cellular automata (QCA)[28-30] operations and strong quantum correlations in Fermi-Hubbard systems.[5]

To achieve precise qubit manipulation and quantum simulation in devices made from TQDs usually requires each TQD to be operated in the few-electron, weakly environmentally coupled regime. An inherent technical difficulty in the experimental control and manipulation of such a few-electron TQD is the precise detection of the charge state transitions in the TQD by measurements of the electric current passing through the TQD, because the current at these charge state transitions is normally extremely small (< 1 pA) and is often below the detection limit of the state-of-the-art electronics or because the transitions between these electron charge states involve charge rearrangements within the three QDs only and thus no current would actually pass through the TQD. Both quantum point contacts (QPCs) and QDs have been used as charge sensors to overcome the problem encountered in the detection of charge state transitions in few-electron, weakly environmentally coupled QD devices, and experiments have shown that a QD charge sensor has advantages in achieving better measurement sensitivity and signal-to-noise ratio due to significant reductions in the screening and lifetime broadening of the quantum states in the QD charge sensor[31,32]. Formations of QPs have been observed in TQD devices made in GaAs/AlGaAs two-dimensional electron gas (2DEG) with the help of integrated QPC charge sensors[16-19]. However, a clear observation of all the eight possible charge states and their accompanied eight triple points and thus a vivid visualization of a full, complex charge stability diagram have not been reported for a TQD made from a nanowire, which is most likely due to the lack of integration of a highly sensitive charge sensor. Several efforts have been undertaken to integrate capacitively-coupled charge sensors in nanowire-based QD systems. Detection of charge states in nanowire-based QDs via integrated QPC charge sensor defined in a GaAs/AlGaAs 2DEG exactly below the nanowire QDs have been realized in early attempts[33,34]. More recently, with the integration of a nanowire-based QD charge sensor to nanowire QDs, charge state transitions in the nanowire QDs have been successfully detected[35-40]. To study the dynamics of degenerate charge configurations at QPs and QD-based QCA operations, and to reveal the full complexity of the charge stability diagram of a nanowire-based TQD, it would be inevitable to integrate a sensitive charge sensor to a TQD in the nanowire platform.

In this article, we report on the realization of a TQD with an integrated QD charge sensor from a single InAs nanowire via a fine finger-gate technique[35-40]. Here, we consider integration of a charge sensor to a TQD on a single InAs nanowire to simplify the device fabrication by eliminating the need for precise placement of nanowires via micromanipulation[41] or pre-patterning on a substrate[42]. We also employ a thin metal wire to enhance the capacitive coupling of the charge sensor to the TQD in order to increase the sensing sensitivity of the internal charge configuration rearrangements in the TQD. The integrated TQD device is characterized by simultaneous measurements of the direct transport signal through the TQD and the detection signal through the QD sensor, and an excellent agreement between these two types of measurements is achieved. The charge stability diagram of the TQD is measured and a capacitance network model[17-19,43,44] that reproduces the measured charge stability diagram is developed. With the help of the highly sensitive charge sensor, we map out the various charge states formed in an energetically degenerated region, where energy levels of all the three QDs are close to be on resonance with the Fermi level of the source and drain reservoirs. The results are presented in a three-dimensional (3D) gate-voltage space via a series of parallel two-dimensional slices through the 3D space. Distinct properties of the TQD, including the formations of all the eight charge states and their accompanied eight triple points in an energetically degenerate region, and of QPs and QD-based QCA cells, are observed and analyzed.

## 2. Results and Discussion
### 2.1 Device architecture and fabrication

The InAs nanowires employed in this work were grown on Si(111) substrates via molecular-beam epitaxy (MBE) using Ag catalysts[45]. Figure 1(a) shows a scanning electron microscope (SEM) image of a representative fabricated device, in which seven top finger gates defining the TQD labeled as G1 to G7, three top finger gates defining the QD sensor labeled as g1 to g3, and the coupling metal wire labeled as CW are displayed. The drain contacts for the TQD and the QD sensor are labeled by D and d, respectively, and the shared source contact is labeled by S. Figure 1(b) shows a schematic 3D view of the device. For device fabrication, the InAs nanowires were transferred by a dry method onto a heavily p-doped Si substrate, covered with a 300-nm-thick layer of $SiO_2$, with predefined metal markers on top. After InAs nanowires with a diameter of ~30 nm and a pure wurtzite crystalline phase were selected and located with respect to the markers, the Ti/Au (5 nm/90 nm in thickness) source and drain contacts were fabricated by standard electron-beam lithography (EBL), metal deposition via electron-beam evaporation (EBE) and lift-off. In order to remove nanowire surface oxides and achieve surface passivation, the contact regions of the nanowires were chemically etched in a $(NH_4)_2S_x$ solution, right before the metal deposition. Subsequently, after another step of EBL, the nanowires were covered by a 10-nm-thick $HfO_2$ layer via atomic layer deposition (ALD) and lift-off. Finally, the Ti/Au (5 nm/25 nm in thickness) top finger gates and coupling metal wire were fabricated by a combined step of EBL, EBE, and lift-off. These top finger gates and the coupling wire had a width of ~20 nm and had formed the arrays with a pitch of ~60 nm. Below, we present our measurements for a device similar to the one shown in Figure 1(a).

**2.2 Capacitive coupling between the QD sensor and the TQD**

The QD sensor and the TQD are capacitively coupled via a coupling metal wire CW [cf. Figure 1(a)]. To investigate the capacitive coupling strength between the sensor QD and the TQD, we first examined a capacitively coupled, parallel DQD formed from a single QD (target QD) in the TQD and the sensor QD by simultaneous measurements of source-drain currents of the target QD ($I_{SD}$) and the sensor QD ($I_{Sd}$). The target QD was defined by using gates G2 and G4 as barrier gates (with gate voltages $V_{G2} = -0.75$ V and $V_{G4} = -0.81$ V) and G3 as a plunger gate. The QD sensor was defined by using gates g1 and g3 as barrier gates (with gate voltages $V_{g1} = -0.85$ V and $V_{g3} = -1.1$ V) and g2 as a plunger gate. Figure 2(a) shows a plot of the superimposed current of the target QD and the sensor QD, $I = I_{SD} + I_{Sd}$, as a function of gate voltages $V_{G3}$ and $V_{g2}$ at bias voltage $V_S = 50$ μV, in which a hexagonal pattern, i.e., the characteristic of a coupled DQD system, is displayed. The finite slopes of the current lines indicate that each QD is capacitively coupled to both plunger gates G3 and g2. Also visible are the distinct shifts in the current lines which occur whenever the current lines belonging to different QDs intersect. Figure 2(b) shows a plot of the current $I_{Sd}$ through the sensor QD only, where the distinct shifts (marked by white dashed circles) of the current lines are more clearly seen. These two QDs together form a parallel DQD with a finite inter-dot capacitive coupling, as we mentioned above, and with no tunnel coupling. The inter-dot capacitance $C_I$ can be determined[43] by the ratio of the current line shift $\Delta V_{G3}^m$ ($\Delta V_{g2}^m$) to the spacing $\Delta V_{G3}$ ($\Delta V_{g2}$), as indicated in Figure 2(a), and the total capacitance of the single QD $C_{\Sigma g2}$ ($C_{\Sigma G3}$):

$$C_I = \frac{\Delta V_{G3}^m}{\Delta V_{G3}} C_{\Sigma g2} = \frac{\Delta V_{g2}^m}{\Delta V_{g2}} C_{\Sigma G3}. \tag{1}$$

Substituting the values of $\Delta V_{G3}^m/\Delta V_{G3} \sim 0.049$ and $\Delta V_{g2}^m/\Delta V_{g2} \sim 0.042$ [extracted from the measurements shown in Figure 2(a)] and the values of $C_{\Sigma g2} \sim 35$ aF and $C_{\Sigma G3} \sim 39$ aF [extracted from the single QD Coulomb diamond measurements, see Part I in the Supplementary Information] into Eq. (1), we obtain $C_I \sim 1.7$ aF. With this considerably large inter-dot capacitance, we can infer that the capacitive coupling between the QD sensor and the TQD is strong enough for the QD sensor to detect the charge state transitions in the TQD. To further confirm this, we also made simultaneous charge detection and direct transport current measurements of the charge state transitions in a DQD defined within the TQD structure and obtained an excellent agreement between the two types of measurements (see Part II in the Supplementary Information for more details).

**2.3 Detection of the charge state transitions in the TQD**

Having successfully demonstrated an excellent detection sensitivity of the QD sensor, we now move to simultaneous charge detection and direct transport current measurements of the TQD defined as schematically shown in Figure 1(c). Here, the TQD is defined by using gates G1, G3, G4 and G6 as barrier gates and gates G2, G5 as plunger gates. The three defined QDs are represented by red ovals and labeled as QD1, QD2, and QD3 as shown in Figure 1(c). Here we should note that QD2 is defined only by gates G3 and G4. Figure 3(a) shows the source-drain current $I_{SD}$ obtained by direct transport measurements through the TQD as a function of gate voltages $V_{G3}$ and $V_{G4}$ at $V_S = 70$ μV. Three groups of parallel current lines, as marked by dashed lines with three different colors, can be identified in Figure 3(a). The appearance of these finite current lines can be attributed to co-tunneling processes when energy levels of any one of the three QDs are on resonance with the Fermi level of the source and drain reservoirs. The slope of the current lines mainly depends on the coupling strengths between the corresponding QD and the swept gates of G3 and G4, which are determined by the relative distances from the corresponding QD to the two swept gates. Accordingly, the current lines marked by almost vertical green dashed lines correspond to resonant transport through energy levels of QD1 in the TQD, since QD1 couples strongly to gate G3 but weakly to gate G4. Similarly, the current lines marked by almost horizontal blue dashed lines correspond to resonant transport through energy levels of QD3 in the TQD. The current lines with a slope of $dV_{G4}/dV_{G3} \sim -1$ as marked by inclined white dashed lines correspond to resonant transport through energy levels of QD2 in the TQD, since QD2 couples almost equally to the two swept gates of G3 and G4.

Figure 3(b) shows the charge stability diagram of the TQD revealed via the QD sensor by operating the QD sensor at a steep slope of a Coulomb current peak (i.e., by setting $V_{g2}=$ 0.108 V at the declining slope of a Coulomb current peak). In detail, the current through the QD sensor $I_{Sd}$ in response to the modulation of the two swept gate voltages is measured in the same range of $V_{G3}$ and $V_{G4}$ as in Figure 3(a) and the transconductance $dI_{Sd}/dV_{G4}$ of the QD sensor is displayed in Figure 3(b). The stability diagram in Figure 3(b) features bright transition lines with three different slopes (i.e., almost vertical, inclined, and almost horizontal lines) as expected for a TQD, which is in good agreement with the charge stability diagram obtained by the direct transport current measurements shown in Figure 3(a). The detection signals for the charge state transitions in the three individual QDs of the TQD appear to be bright lines of positive transconductance ($dI_{Sd}/dV_{G4}$). This is because adding one electron from the reservoirs to the TQD with increasing $V_{G4}$ leads to a positive transconductance $dI_{Sd}/dV_{G4}$ as the QD sensor is set at the declining slope of a Coulomb current peak. The main features of the experimentally measured charge stability diagrams shown in Figures 3(a) and 3(b) can be well reproduced by simulation using a capacitance network model. Figure 3(c) displays the equivalent circuit diagram for our TQD and Figure 3(d) shows the simulated stability diagram, which agrees well with the experimental results (see Part III in the Supplementary Information).

**2.4 Complex charge stability diagram, QPs and QCA processes**

To explore the rich charge state configuration physics of the TQD in a region where the energy levels of all the three QDs are on resonance with each other, the charge stability diagram is measured via the QD charge sensor as a function of three gate voltages $V_{G2}$, $V_{G3}$ and $V_{G4}$ and is presented by a series of parallel two-dimensional charge stability diagrams sliced through the 3D gate-voltage space. Figures 4(a)-4(e) show the evolution of the charge stability diagrams of the TQD as in Figure 3 with changes in gate voltage $V_{G2}$, which primarily controls the electrostatic potential of QD1. Here, the QD sensor is operated by setting $V_{g2}$= 0.108 V [the same as in Figure 3(b)] and the transconductance $dI_{Sd}/dV_{G4}$ of the QD sensor is measured as a function of $V_{G3}$ and $V_{G4}$ at $V_S$ = 70 μV and at different $V_{G2}$. Due to finite electrostatic inter-dot couplings, all intersections between charge state transition lines from different groups are avoided, resulting in pairs of triple points connected via charge-conserved charge state transition lines in which charge transfers from one QD to another without a change in the total number of charges in the TQD. These charge-conserved charge state transition lines appear to be dark lines instead of bright lines. This is because increasing $V_{G4}$ across a dark charge-conserved charge state transition line will lead to a transfer of an electron charge from QD1 to QD2 or from QD2 to QD3, giving a negative transconductance $dI_{Sd}/dV_{G4}$, since in our device geometry the coupling strength between the QD sensor and QD1 (QD2) is stronger than that between the QD sensor and QD2 (QD3). The lengths of these dark lines are proportional to the electrostatic inter-dot coupling strengths between the two corresponding QDs. It can be discerned from Figure 3(b) that the electrostatic inter-dot coupling strength between neighboring QDs is much stronger than that between QD1 and QD3, which is consistent with the device geometry as QD1 and QD3 are separated by QD2.

In an energetically degenerate region where energy levels of all the three QDs are on or closely on resonance with each other, charge transition lines of all the three QDs meet, resulting in $2^3$=8 possible charge states of the TQD. All these eight charge states are observed in the sensor-detected charge stability diagram shown in Figure 4(c). In accompany with these eight charge states, there are eight triple points, i.e., four pairs of triple points with each pair being connected by a dark charge-conserved charge state transition line. For a clear description of this result, we have simply assigned the charge state of the TQD in the lower left region of Figure 4(c) as (0, 0, 0) and denote the observed stable charge states in the other regions as ($N_1$, $N_2$, $N_3$), where $N_i$=0 or 1 corresponds to an additional electron number in QDi. We would like to emphasize clearly again that our measured TQD is in the few-electron regime but not in the single electron regime and that exact electron occupation numbers in the TQD is not known. Thus, here and after in the article, $N_i$ should be considered as an extra electron occupation number in QDi measured relatively to that in the lower-left gate-voltage region of Figure 4(c). The observed eight triple points, which are the degenerate points of three different charge states, are marked by white dots and labeled as A-H in Figure 4(c). Here all the eight possible charge states and their accompanied eight triple points are observed in the measured charge stability diagram of the TQD, which is achieved with recent advances in our semiconductor nanowire QD device fabrication technology.

We now demonstrate tuning of the TQD to QPs. As shown in Figure 4(c) to Figure 4(e), with increasing gate voltage $V_{G2}$ from $V_{G2} = 0.194$ V to $V_{G2} = 0.1947$ V, triple points A and B are getting closer and at the same time triple points D and F are moving closer [see Figure 4(d)]. With further increasing $V_{G2}$ to $V_{G2} = 0.195$ V, the two triple points A and B finally merge to form a QP as marked by a white dot $Q_{AB}$ [see Figure 4(e)]. At the QP $Q_{AB}$, the four charge states (0, 0, 0), (1, 0, 0), (0, 1, 0) and (0, 0, 1) are energetically degenerate and an electron can sequentially tunnel from the source reservoir to QD1, QD2, QD3 and then to the drain reservoir. At the same time, the triple points D and F also merge to form a QP as marked by a white dot $Q_{DF}$ in Figure 4(e). At this QP, the four charge states (1, 0, 0), (1, 1, 0), (1, 0, 1) and (0, 1, 0) are energetically degenerate. Here, the electron tunneling sequence differs from that at QP $Q_{AB}$ and electrons are transferred through the TQD in the following manner. Starting with charge state (1, 0, 0), an electron tunnels from the drain reservoir to QD3 to reach charge state (1, 0, 1) and then tunnels from QD3 to QD2 to reach charge state (1, 1, 0). Subsequently, the electron in QD1 tunnels to the source reservoir to leave the TQD at charge state (0, 1, 0). Finally, the electron in QD2 tunnels to QD1 to restore the initial charge state (1, 0, 0). In a similar way, as $V_{G2}$ is decreased to $V_{G2} = 0.193$ V and then to $V_{G2} = 0.1926$ V, as shown from Figure 4(c) to Figure 4(a), triple points G and H and triple points C and E are getting closer and then merge to form QPs labeled as $Q_{GH}$ and $Q_{CE}$. At QP $Q_{GH}$, like at QP $Q_{AB}$, a hole can sequentially tunnel through the TQD via the transition sequence of (1, 1, 1) ↔ (1, 1, 0) ↔ (1, 0, 1) ↔ (0, 1, 1) ↔ (1, 1, 1). At QP $Q_{CE}$, a similar process as at QP $Q_{DF}$ occurs via the transition sequence of (0, 1, 1) ↔ (0, 1, 0) ↔ (0, 0, 1) ↔ (1, 0, 1) ↔ (0, 1, 1).

Finally, we demonstrate that the nanowire TQD can also be operated as a QCA, in which adding an electron from a reservoir to one outer QD or removing an electron in an outer QD to a reservoir will automatically lead to an electron transfer between the other two QDs. As shown in Figure 4(c), when the nanowire TQD is tuned across the charge state transition line in between triple points C and F from charge state (0, 1, 0) to charge state (1, 0, 1), adding an electron from source reservoir to QD1 will automatically lead to moving the electron in QD2 to QD3. In a reverse process, i.e., when the TQD is tuned from charge state (1, 0, 1) to charge state (0, 1, 0), removing an electron from QD1 or QD3 to a reservoir will automatically make the other electron favorably stay in QD2. These processes, harnessing special conditions of Coulomb interaction in a TQD, may lead to applications in building up a QD-based QCA circuit.

## 3. Conclusions

We have realized the integration of a highly sensitive QD charge sensor to a serial TQD in a single InAs nanowire via a fine finger-gate technique and studied the complex charge stability diagrams of the nanowire TQD. With the help of the QD charge sensor, the charge-conserved charge state transitions in the TQD, in which there is no change in the total number of electrons in the TQD and thus no transport current is detectable, are observed, allowing a clear identification of all the eight possible charge states and all the eight possible triple points in an energetically degenerated region, where all the three QDs are on or nearly on resonance with each other. A series of parallel two-dimensional charge stability diagram slices through the 3D gate voltage space are measured and a detailed examination of the charge state evolution near the energetically degenerated region allows us to detect QPs, where a fourfold charge state degeneracy occurs, and to operate the TQD as a QCA. This integrated nanowire multiple QD system is expected to provide a versatile platform for developments of semiconductor-nanowire based quantum logic processors and quantum simulators. However, to demonstrate the potential of the TQD or nanowire-based multiple-QD qubits towards this direction, a real-time single-shot readout measurement of the charge is required. Unfortunately, the bandwidth of the integrated QD charge sensor in our currently studied device is limited by our present measurement setup to less than a few kHz, making such a real-time single-shot readout measurement impossible. Our next step is to develop a radio-frequency reflectometry technique. We anticipate that by combining our current advanced device fabrication technique with a radio-frequency reflectometry technique, fast high fidelity single-shot readout of semiconductor-nanowire QD qubits would be achieved.

## 4. Methods

*Measurement set-up*: All transport measurements presented here were performed in a He$^3$/He$^4$ dilution refrigerator at a base temperature of ~20 mK. Figure 1(c) shows a schematic cross-sectional view of the device and measurement circuit setup. Both the TQD and the QD sensor were measured in a two-terminal DC setup with a bias voltage $V_S$ applied to the shared source contact and the drain contacts kept grounded. The Si substrate and the SiO$_2$ layer were employed as a global back gate and a gate dielectric, respectively. Throughout the measurements, the global back gate voltage $V_{BG}$ was fixed at 6.3 V to keep the nanowires in a n-type conduction state. The transfer characteristics of individual finger gates were measured and the current pinch-off voltages of all the finger gates were found in the range of −1.1 V to −0.7 V. The uninvolved gate G7 was grounded throughout the measurements.

## Acknowledgements

This work is supported by the National Natural Science Foundation of China (Grant Nos. 92165208, 91221202, 91421303, 11874071, 92065106 and 61974138), the Ministry of Science and Technology of China through the National Key Research and Development Program of China (Grant Nos. 2017YFA0303304 and 2016YFA0300601), and the Beijing Academy of Quantum Information Sciences (No. Y18G22). DP also acknowledges the support from Youth Innovation Promotion Association, Chinese Academy of Sciences (Nos. 2017156 and Y2021043).

## Competing interests

The authors declare no competing interests.

## Data availability

The data that support the findings of this study are available within the article and Supplementary Information, and from the corresponding author upon reasonable request.

**References**

1. Loss, D. & D. P. DiVincenzo, D. P. Quantum computation with quantum dots. Phys. Rev. A **57**, 120-126 (1998).
2. Burkard, G., Ladd, T. D., Nichol, J. M., Pan, A. & Petta, J. R. Semiconductor Spin Qubits. Preprint at https://doi.org/10.48550/arXiv.2112.08863 (2022).
3. Georgescu, I. M., Ashhab, S. & Nori, F. Quantum simulation. Rev. Mod. Phys. **86**, 153-185 (2014).
4. Byrnes, T., Kim, N. Y., Kusudo, K. & Yamamoto, Y. Quantum simulation of Fermi-Hubbard models in semiconductor quantum-dot arrays. Phys. Rev. B **78**, 075320 (2008).
5. Hensgens, T., Fujita, T., Janssen, L., Li, X., Van Diepen, C. J., Reichl, C., Wegscheider, W., Das Sarma, S. & Vandersypen, L. M. K. Quantum simulation of a Fermi–Hubbard model using a semiconductor quantum dot array. Nature **548**, 70-73 (2017).
6. Van Diepen, C. J., Hsiao, T. K., Mukhopadhyay, U., Reichl, C., Wegscheider, W. & Vandersypen, L. M. K. Quantum Simulation of Antiferromagnetic Heisenberg Chain with Gate-Defined Quantum Dots. Phys. Rev. X **11**, 041025 (2021).
7. Fasth, C., Fuhrer, A., Samuelson, L., Golovach, V. N. & Loss, D. Direct Measurement of the Spin-Orbit Interaction in a Two-Electron InAs Nanowire Quantum Dot. Phys. Rev. Lett. **98**, 266801 (2007).
8. Csonka, S., Hofstetter, L., Freitag, F., Oberholzer, S., Schönenberger, C., Jespersen, T. S., Aagesen, M. & Nygård, J. Giant Fluctuations and Gate Control of the *g*-Factor in InAs Nanowire Quantum Dots. Nano Lett. **8**, 3932-3935 (2008).
9. Nilsson, H. A., Caroff, P., Thelander, C., Larsson, M., Wagner, J. B., Wernersson, L. E., Samuelson, L. & Xu, H. Q. Giant, Level-Dependent g Factors in InSb Nanowire Quantum Dots. Nano Lett. **9**, 3151-3156 (2009).
10. Wang, J.-Y., Huang, S., Lei, Z., Pan, D., Zhao, J. & Xu, H. Q. Measurements of the spin-orbit interaction and Landé g factor in a pure-phase InAs nanowire double quantum dot in the Pauli spin-blockade regime. Appl. Phys. Lett. **109**, 053106 (2016).
11. Wang, J.-Y., Huang, G.-Y., Huang, S., Xue, J., Pan, D., Zhao, J. & Xu, H. Q. Anisotropic Pauli Spin-Blockade Effect and Spin–Orbit Interaction Field in an InAs Nanowire Double Quantum Dot. Nano Lett. **18**, 4741-4747 (2018).
12. Iorio, A., Rocci, M., Bours, L., Carrega, M., Zannier, V., Sorba, L., Roddaro, S., Giazotto, F. & Strambini, E. Vectorial Control of the Spin–Orbit Interaction in Suspended InAs Nanowires. Nano Lett. **19**, 652-657 (2019).

13. Mu, J., Huang, S., Wang, J.-Y., Huang, G.-Y., Wang, X. & Xu, H. Q. Measurements of anisotropic g-factors for electrons in InSb nanowire quantum dots. Nanotechnology **32**, 020002 (2020).

14. Nadj-Perge, S., Frolov, S. M., Bakkers, E. P. A. M. & Kouwenhoven, L. P. Spin–orbit qubit in a semiconductor nanowire. Nature **468**, 1084-1087 (2010).

15. Nadj-Perge, S., Pribiag, V. S., van den Berg, J. W. G., Zuo, K., Plissard, S. R., Bakkers, E. P. A. M., Frolov, S. M. & Kouwenhoven, L. P., Fast Spin-Orbit Qubit in an Indium Antimonide Nanowire. Phys. Rev. Lett. **108**, 166801 (2013).

16. Gaudreau, L., Studenikin, S. A., Sachrajda, A. S., Zawadzki, P., Kam, A., Lapointe, J., Korkusinski, M. & Hawrylak, P. Stability Diagram of a Few-Electron Triple Dot. Phys. Rev. Lett. **97**, 036807 (2006).

17. Schröer, D., Greentree, A. D., Gaudreau, L., Eberl, K., Hollenberg, L. C. L., Kotthaus, J. P. & Ludwig, S. Electrostatically defined serial triple quantum dot charged with few electrons. Phys. Rev. B **76**, 075306 (2007).

18. Rogge, M. C. & Haug, R. J. The three dimensionality of triple quantum dot stability diagrams. New J. Phys. **11**, 113037 (2009).

19. Granger, G., Gaudreau, L., Kam, A., Pioro-Ladrière, M., Studenikin, S. A., Wasilewski, Z. R., Zawadzki, P. & Sachrajda, A. S. Three-dimensional transport diagram of a triple quantum dot. Phys. Rev. B **82**, 075304 (2010).

20. Wang, J.-Y., Huang, S., Huang, G. Y., Pan, D., Zhao, J. & Xu, H. Q. Coherent Transport in a Linear Triple Quantum Dot Made from a Pure-Phase InAs Nanowire. Nano Lett. **17**, 4158-4164 (2017).

21. Hiraoka, S., Horibe, K., Ishihara, R., Oda, S. & Kodera, T. Physically defined silicon triple quantum dots charged with few electrons in metal-oxide-semiconductor structures. Appl. Phys. Lett. **117**, 074001 (2020).

22. Taylor, J. M., Srinivasa, V. & Medford, J. Electrically Protected Resonant Exchange Qubits in Triple Quantum Dots. Phys. Rev. Lett. **111**, 050502 (2013).

23. Medford, J., Beil, J., Taylor, J. M., Rashba, E. I., Lu, H., Gossard, A. C. & Marcus, C. M. Quantum-Dot-Based Resonant Exchange Qubit. Phys. Rev. Lett. **111**, 050501 (2013).

24. Eng, K., Ladd, T. D., Smith, A., Borselli, M. G., Kiselev, A. A., Fong, B. H., Holabird, K. S., Hazard, T. M., Huang, B., Deelman, P. W., Milosavljevic, I., Schmitz, A. E., Ross,

R. S., Gyure, M. F. & Hunter, A. T. Isotopically enhanced triple-quantum-dot qubit. Sci. Adv. **1**, e1500214 (2015).

25. Russ, M., Petta, J. R. & Burkard, G. Quadrupolar Exchange-Only Spin Qubit. Phys. Rev. Lett. **121**, 177701 (2018).

26. Qiao, H., Kandel, Y. P., Manikandan, S. K., Jordan, A. N., Fallahi, S., Gardner, G. C., Manfra M. J. & Nichol, J. M. Conditional teleportation of quantum-dot spin states. Nat. Commun. **11**, 3022 (2020).

27. Kojima, Y., Nakajima, T., Noiri, A., Yoneda, J., Otsuka, T., Takeda, K., Takeda, K., Li, S., Bartlett, S. D., Ludwig, A., Wieck, A. D. & Tarucha, S. Probabilistic teleportation of a quantum dot spin qubit. npj Quantum Inf. **7**, 68 (2021).

28. Lent, C. S., Tougaw, P. D., Porod, W. & Bernstein, G. H. Quantum cellular automata. Nanotechnology **4**, 49-57 (1993).

29. Amlani, I., Alexei, O. O., Toth, G., Gary, H. B., Craig, S. L. & Gregory, L. S. Digital Logic Gate Using Quantum-Dot Cellular Automata. Science **284**, 289-291 (1999).

30. Tóth, G. & Lent, C. S. Quantum computing with quantum-dot cellular automata. Phys. Rev. A **63**, 052315 (2001).

31. Barthel, C., Kjærgaard, M., Medford, J., Stopa, M., Marcus, C. M., Hanson, M. P. & Gossard, A. C. Fast sensing of double-dot charge arrangement and spin state with a radio-frequency sensor quantum dot. Phys. Rev. B, **81**, 161308 (2010).

32. Rössler, C., Krähenmann, T., Baer, S., Ihn, T., Ensslin, K., Reichl, C. & Wegscheider, W. Tunable charge detectors for semiconductor quantum circuits. New J. Phys. **15**, 033011 (2013).

33. Wallin, D., Fuhrer, A., Fröberg, L. E., Samuelson, L., Xu, H. Q., Hofling, S. & Forchel, A. Detection of charge states in nanowire quantum dots using a quantum point contact. Appl. Phys. Lett. **90**, 172112 (2007).

34. Shorubalko, I., Leturcq, R., Pfund, A., Tyndall, D., Krischek, R., Schön, S. & Ensslin, K. Self-Aligned Charge Read-Out for InAs Nanowire Quantum Dots. Nano Lett. **8**, 382-385 (2008).

35. Hu, Y., Churchill, H. O. H., Reilly, D. J., Xiang, J., Lieber, C. M. & Marcus, C. M. A Ge/Si heterostructure nanowire-based double quantum dot with integrated charge sensor. Nat. Nanotechnol. **2**, 622-625 (2007).

36. Zhou, X. & Ishibashi, K. Single charge detection in capacitively coupled integrated single electron transistors based on single-walled carbon nanotubes. Appl. Phys. Lett.

**101**, 123506 (2012).

37. Hu, Y., Kuemmeth, F., Lieber, C. M. & Marcus, C. M. Hole spin relaxation in Ge–Si core–shell nanowire qubits. Nat. Nanotechnol. **7**, 47-50 (2012)

39. Li, W., Mu, J., Huang, S., Pan, D., Zhao, J. & Xu, H. Q. Detection of charge states of an InAs nanowire triple quantum dot with an integrated nanowire charge sensor. Appl. Phys. Lett. **117**, 262102 (2020).

40. Wang, X., Huang, S., Wang, J.-Y., Pan, D., Zhao, J. & Xu, H. Q. A charge sensor integration to tunable double quantum dots on two neighboring InAs nanowires. Nanoscale **13**, 1048-1054 (2021).

41. Flöhr, K., Liebmann, M., Sladek, K., Günel, H. Y., Frielinghaus, R., Haas, F., Meyer, C., Hardtdegen, H., Schäpers, T., Grützmacher, D. & Morgenstern, M. Manipulating InAs nanowires with submicrometer precision. Rev. Sci. Instrum. **82**, 113705 (2011).

42. Lim, J. K., Lee, B. Y., Pedano, M. L., Senesi, A. J., Jang, J.-W., Shim, W., Hong, S. & Mirkin, C. A. Alignment Strategies for the Assembly of Nanowires with Submicron Diameters. Small **6**, 1736-1740 (2010).

43. van der Wiel, W. G., De Franceschi, S., Elzerman, J. M., Fujisawa, T., Tarucha, S. & Kouwenhoven, L. P., Electron transport through double quantum dots. Rev. Mod. Phys. **75**, 1-22 (2002).

44. Hanson, R., Kouwenhoven, L. P., Petta, J. R., Tarucha, S. & Vandersypen, L. M. K. Spins in few-electron quantum dots. Rev. Mod. Phys. **79**, 1217-1265 (2007).

45. Pan, D., Fu, M., Yu, X., Wang, X., Zhu, L., Nie, S., Wang, S., Chen, Q., Xiong, P., von Molnár, S. & Zhao, J. Controlled Synthesis of Phase-Pure InAs Nanowires on Si(111) by Diminishing the Diameter to 10 nm. Nano Lett. **14**, 1214-1220 (2014).

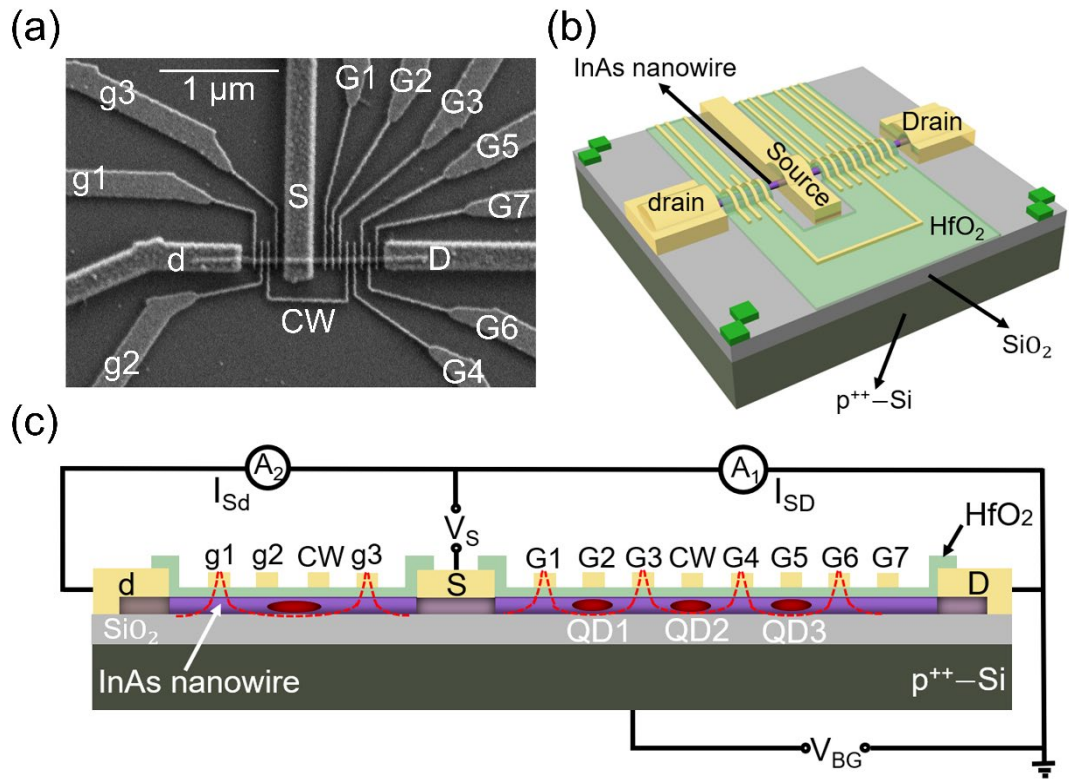

**Figure 1.** (a) SEM image of a QD charge sensor integrated TQD device built from a MBE-grown single InAs nanowire via a fine top finger-gate technique. Finger gates G1-G7 are fabricated to define the TQD and finger gates g1-g3 are fabricated to define the QD charge sensor. The charge sensor is coupled to the TQD via a thin metal wire which is labeled as CW in the image. The finger gates and the coupling wire have a width of ~20 nm and a pitch of ~60 nm. (b) Schematic view of the three-dimensional structure of the device. (c) Cross-sectional schematic view of the device and measurement circuit setup. The locations of the sensor QD and of the three QDs formed in the TQD are marked by red ovals.

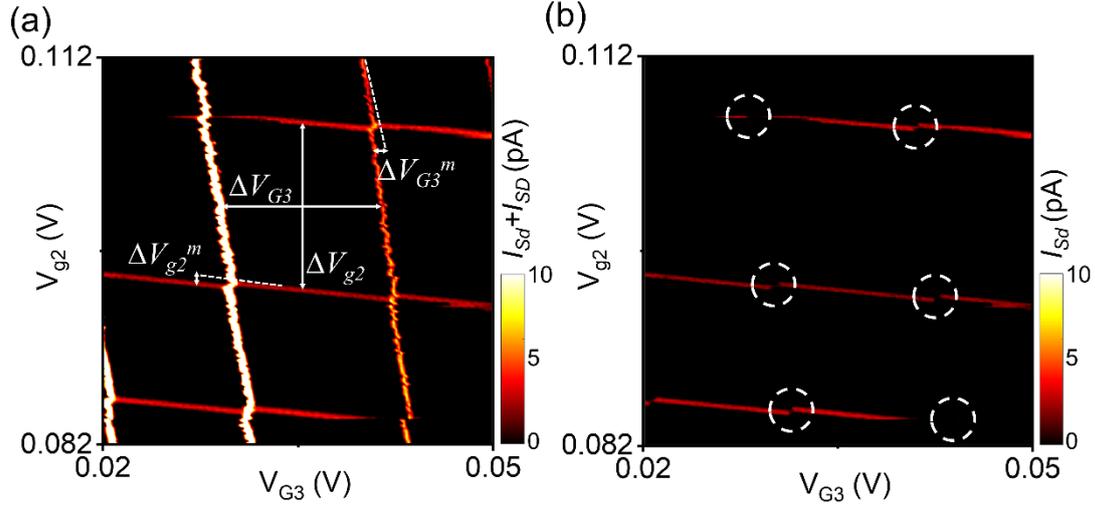

**Figure 2.** (a) Superimposed current of a single QD defined in the TQD (target QD) and the QD sensor, $I = I_{SD} + I_{Sd}$, as a function of gate voltages $V_{G3}$ and $V_{g2}$ measured at $V_S = 50~\mu V$, where the current line shifts $\Delta V_{G3}^m$ and $\Delta V_{g2}^m$ and the current line spacings $\Delta V_{G3}$ and $\Delta V_{g2}$ are shown. Here, $I_{SD}$ is the current passing through the target QD and $I_{Sd}$ is the current passing through the QD sensor. (b) The same as (a) but only the current $I_{Sd}$ of the QD sensor is plotted. The distinct shifts of the current $I_{Sd}$ lines are more clearly seen and appear in the regions marked by white dashed circles.

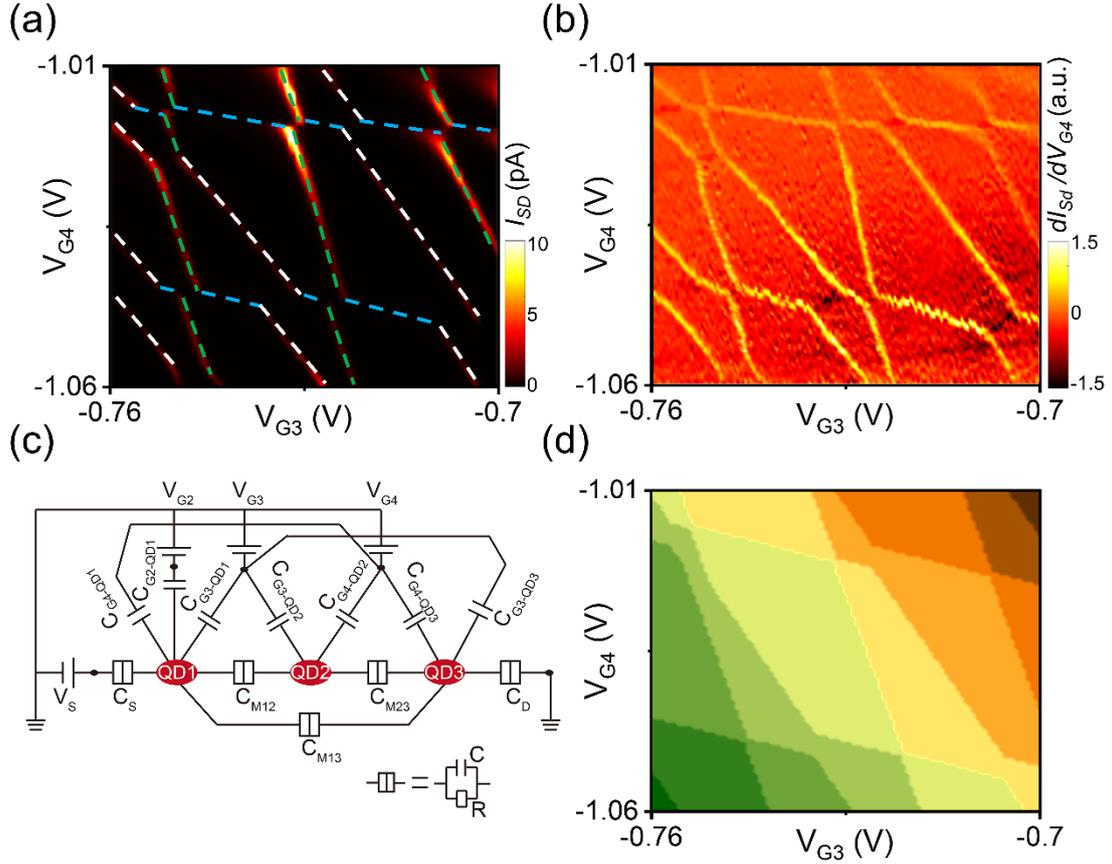

**Figure 3.** (a) Source-drain current $I_{SD}$ of the TQD as a function of gate voltages $V_{G3}$ and $V_{G4}$ at $V_S$ = 70 μV. The TQD is defined and manipulated using gates G1-G6 (with the settings of $V_{G1}$= −0.735 V, $V_{G2}$= 0.191 V, $V_{G5}$= 0.2 V, and $V_{G6}$= −0.91 V). (b) Charge stability diagram of the TQD measured as the transconductance $dI_{Sd}/dV_{G4}$ of the QD sensor as a function of $V_{G3}$ and $V_{G4}$. Here, the results shown in (a) and (b) are measured simultaneously. The dashed lines of different colors mark the current lines originating from resonant transport through energy levels of different individual QDs (see the text for details). (c) Equivalent circuit diagram of the capacitance network model employed to simulate the charge stability diagram of the serial TQD, where the three QDs are marked with red ovals. (d) Simulated charge stability diagram of the TQD based on the capacitance network model.

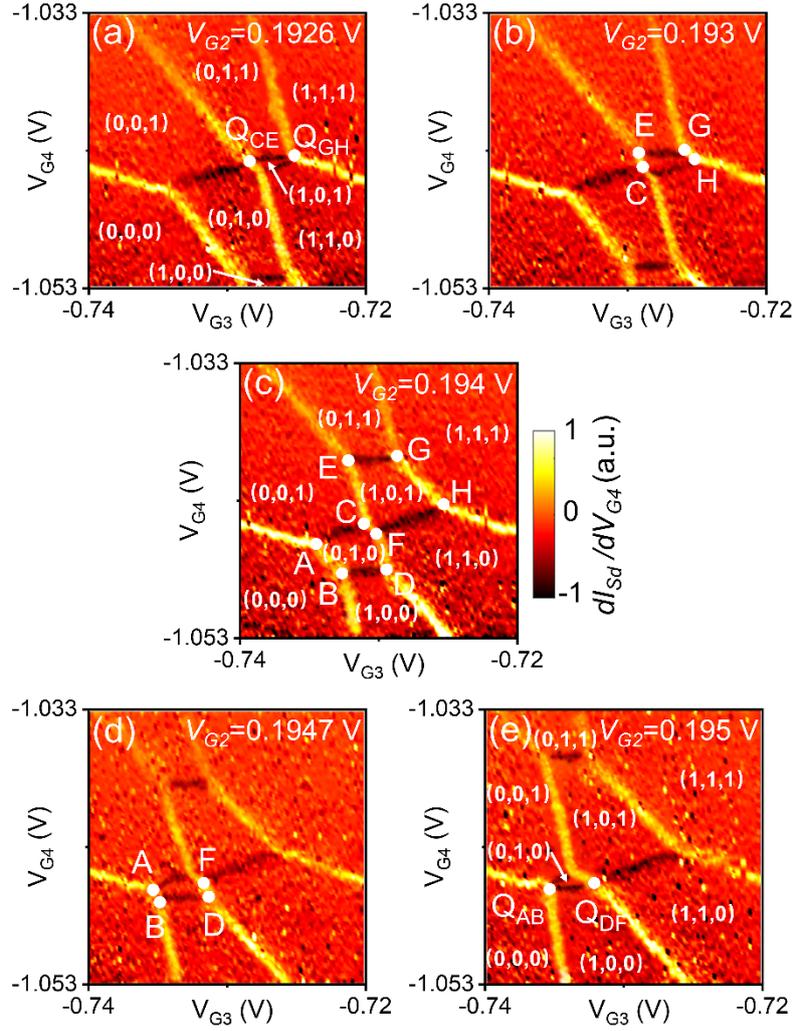

**Figure 4.** (a)-(e) Charge stability diagrams of the TQD measured as the transconductance $dI_{Sd}/dV_{G4}$ of the QD charge sensor at $V_S = 70$ μV as a function of $V_{G3}$ and $V_{G4}$ at different $V_{G2}$. The white dots labeled as A-H in (b), (c) and (d) mark the triple points where the energy levels of two of the three QDs are on resonance with each other and with the Fermi level of the source and drain reservoirs. The white dots labeled as $Q_{AB}$, $Q_{DF}$, $Q_{CE}$ and $Q_{GH}$ in (a) and (e) mark the QPs where the energy levels of all the three QDs are on resonance with each other and with the Fermi level of the reservoirs. Here we note that our measured TQD is in the few-electron regime but with the exact electron number in each QD unknown. However, for clarity, we have assigned the charge state of the TQD in the lower-left region of panels (a), (c) and (e) to (0, 0, 0) and denote the all involved charge states here as ($N_1$, $N_2$, $N_3$), where $N_i$ stands for an electron occupation number in QDi measured relatively to that in the lower-left region of the panels.

# Supplementary Information for
# Charge states, triple points and quadruple points in an InAs nanowire triple quantum dot revealed by an integrated charge sensor

Weijie Li,[1,2] Zhihai Liu,[3] Jingwei Mu,[1,2] Yi Luo,[1,4] Dong Pan,[5] Jianhua Zhao,[5] and H. Q. Xu[1,3,*]

[1] Beijing Key Laboratory of Quantum Devices, Key Laboratory for the Physics and Chemistry of Nanodevices, and School of Electronics, Peking University, Beijing 100871, China

[2] Academy for Advanced Interdisciplinary Studies, Peking University, Beijing 100871, China

[3] Beijing Academy of Quantum Information Sciences, Beijing 100193, China

[4] Institute of Condensed Matter and Material Physics, School of Physics, Peking University, Beijing 100871, China

[5] State Key Laboratory of Superlattices and Microstructures, Institute of Semiconductors, Chinese Academy of Sciences, P.O. Box912, Beijing 100083, China

[*] Author to whom correspondence should be addressed: hqxu@pku.edu.cn

(Dated: January 30, 2023)

## I. Finite bias spectroscopy measurements of single QDs

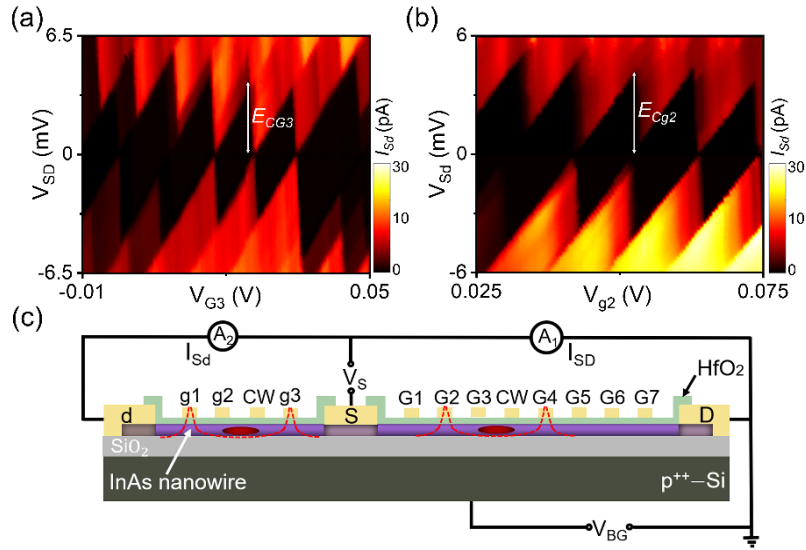

**Figure S1.** (a) Measured source-drain current $I_{SD}$ of a QD (target QD) defined in the region marked by the right red oval in (c) as a function of source-drain bias voltage $V_{SD}$ and gate voltage $V_{G3}$ (charge stability diagram of the target QD). Here the target QD is defined by setting the voltage applied to gates G2 and G4 at $V_{G2} = -0.75$ V and $V_{G4} = -0.81$ V. (b) Measured source-drain current $I_{Sd}$ of the QD sensor defined in the region marked by the left red oval in (c) as a function of source-drain bias voltage $V_{SD}$ and gate voltage $V_{g2}$ (charge stability diagram of the sensor QD). Here the sensor QD is defined by setting the voltage applied to gates g1 and g3 at $V_{g1} = -0.85$ V and $V_{g3} = -1.1$ V. (c) Cross-sectional schematic view of the target QD and the integrated sensor QD. The two red ovals mark the locations of the two QDs.

To investigate the capacitive coupling strength between the quantum dot (QD) sensor and the TQD, we make simultaneous measurements of a capacitively coupled double-QD (DQD) consisting of a single QD (target QD) in the TQD and the QD sensor, as schematically shown in Figure S1(c). Figure S1(a) shows the measured charge stability diagram of the target QD as a function of source-drain bias voltage $V_{SD}$ and voltage $V_{G3}$ applied to plunger gate G3. From the measured Coulomb diamonds, the charging energy of the target QD is estimated to be $E_{CG3} \sim 4$ meV and the total capacitance of the target QD is extracted to be $C_{\Sigma G3} \sim 39$ aF. Figure S1(b) shows the charge stability diagram of the QD sensor as a function of source-drain bias voltage $V_{Sd}$ and voltage $V_{g2}$ applied to plunger gate g2. From the measured Coulomb diamonds, the charging energy of the sensor QD is estimated to be $E_{Cg2} \sim 4.5$ meV and the total capacitance of the sensor QD is extracted to be $C_{\Sigma g2} \sim 35$ aF.

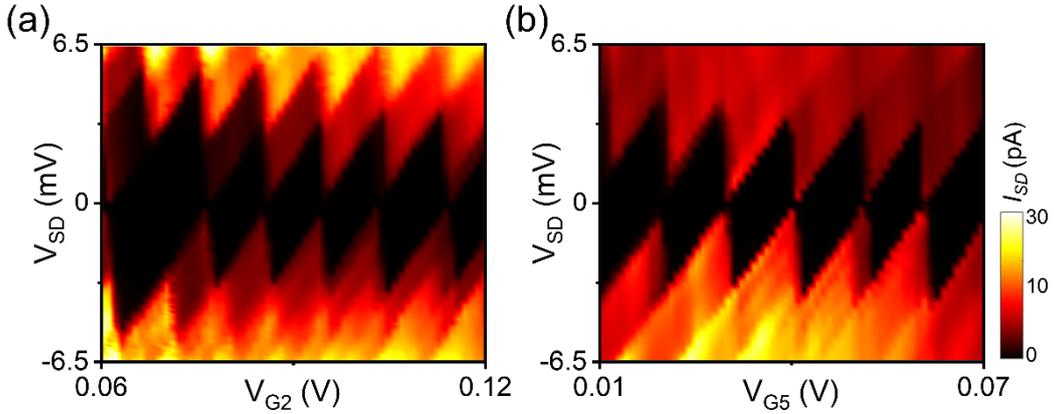

**Figure S2.** (a) Measured source-drain current $I_{SD}$ of QD1 as defined in Figure 1(c) of the main article as a function of source-drain bias voltage $V_{SD}$ and gate voltage $V_{G2}$ (charge stability diagram). Here, QD1 is defined by setting the voltage applied to gates G1 and G3 at $V_{G1} = -0.74$ V and $V_{G3} = -0.8$ V. (b) Measured source-drain current $I_{SD}$ of QD3 as defined in Figure 1(c) of the main article as a function of source-drain bias voltage $V_{SD}$ and gate voltage $V_{G5}$ (charge stability diagram). Here, QD3 is defined by setting the voltage applied to gates G4 and G6 at $V_{G4} = -1.05$ V and $V_{G6} = -0.92$ V.

Figure S2(a) shows the charge stability diagram of a single QD (QD1) in the TQD, as defined in Figure 1(c) of the main article, as a function of source-drain bias voltage $V_{SD}$ and voltage $V_{G2}$ applied to plunger gate G2. Here, QD1 was defined by using gates G1 and G3 as the two outer tunneling barrier gates (with gate voltages fixed at $V_{G1} = -0.74$ V and $V_{G3} = -0.8$ V) and G2 as the plunger gate (with the remaining gates G4 to G7 being grounded). The regular Coulomb diamonds as well as the close points seen at zero $V_{SD}$ between neighboring Coulomb diamonds indicate the formation of a single QD (QD1). The two slopes of the two edges of the Coulomb diamonds depend on the capacitances and can be expressed as $\frac{C_1 - C_{S1} - C_{D1}}{C_1 - C_{S1}}$ and $-\frac{C_1 - C_{S1} - C_{D1}}{C_{S1}}$, where $C_1$, $C_{S1}$ and $C_{D1}$ are the total capacitance, the source capacitance and the drain capacitance of QD1, respectively. Figure S2(b) shows the charge stability diagram of another single

QD (QD3) in the TQD as defined in Figure 1(c) of the main article as a function of source-drain bias voltage $V_{SD}$ and voltage $V_{G5}$ applied to plunger G5. Here, QD3 was defined by using gates G4 and G6 as the two outer tunneling barrier gates (with gate voltages fixed at $V_{G4} = -1.05$ V and $V_{G6} = -0.92$ V) and G5 as the plunger gate (with the remaining gates G1 to G3 and G7 being grounded). The two slopes of the two edges of the Coulomb diamonds shown in Figure S2(b) can be expressed as $\frac{C_3 - C_{S3} - C_{D3}}{C_3 - C_{S3}}$ and $-\frac{C_3 - C_{S3} - C_{D3}}{C_{S3}}$, where $C_3$, $C_{S3}$ and $C_{D3}$ are the total capacitance, the source capacitance and the drain capacitance of QD3, respectively. We can extract the source capacitance $C_{S1}$ of QD1 and the drain capacitance $C_{D3}$ of QD3 from the measured Coulomb diamonds in Figures S2(a) and S(b), and estimate the source and drain capacitances of the TQD defined as schematically shown in Figure 1(c) of the main article to be $C_S \approx C_{S1} \sim 9.80$ aF and $C_D \approx C_{D3} \sim 8.00$ aF (see Table SI).

**II. Simultaneous charge detection and direct transport measurements of a DQD**

To further verify the excellent sensitivity of the QD sensor, we made simultaneous charge detection and direct transport current measurements of a double-QD (DQD) defined in the InAs nanowire as schematically shown in Figure S3(g). The DQD was defined by using gates G1 and G5 as the two outer tunneling barrier gates (with gate voltages fixed at $V_{G1} = -0.75$ V and $V_{G5} = -0.81$ V), G3 as the inter-dot tunneling barrier gate, and G2 and G4 as the two plunger gates. Figures S3(a)-S3(c) show the source-drain current $I_{SD}$ obtained by direct transport current measurements of the DQD at $V_S = 70$ μV as a function of voltages $V_{G2}$ and $V_{G4}$ applied to gates G2 and G4 at three representative inter-dot coupling strengths. At gate voltage $V_{G3} = 0$ V [Figure S3(a)], a large single QD is defined between barrier gates G1 and G5 as there should be no tunneling barrier formed in the nanowire by gate G3. As seen in Fig. S1(a), clear straight diagonal current lines are observed, in agreement with the fact that only a single QD is effectively formed in the nanowire between gates G1 and G5. Pushing the voltage applied to G3 to a negative value of $V_{G3} = -0.65$ V [Figure S3(b)] generates a finite tunneling barrier in the nanowire and a DQD can thus be formed between barrier gates G1 and G5. As shown in Figure S3(b), some measured current lines are seen to be clearly bent. These bents occur when independent energy levels in the two QDs move close in energy. Pairs of kink points with large separations are also observable, indicating that the formed DQD is in the strong inter-dot coupling regime. With tuning $V_{G3}$ to $V_{G3} = -0.88$ V, a high tunneling barrier is generated in the nanowire and well-defined hexagon-shaped patterns are obtained in the measured current $I_{SD}$ as a function of $V_{G2}$ and $V_{G4}$ [Figure S3(c)], indicating that the DQD is in the weak inter-dot coupling regime. Figures S1(d)-S1(f) shows the charge stability diagrams of the TQD detected by the QD sensor with gate voltage $V_{g2}$ set at $V_{g2} = 0.108$ V (i.e., at the declining slope of a Coulomb current peak of the sensor QD), where the transconductance $dI_{Sd}/dV_{G4}$ of the QD sensor is measured in the same gate voltage regions of $V_{G2}$ and $V_{G4}$ as in Figures S3(a)-S3(c).

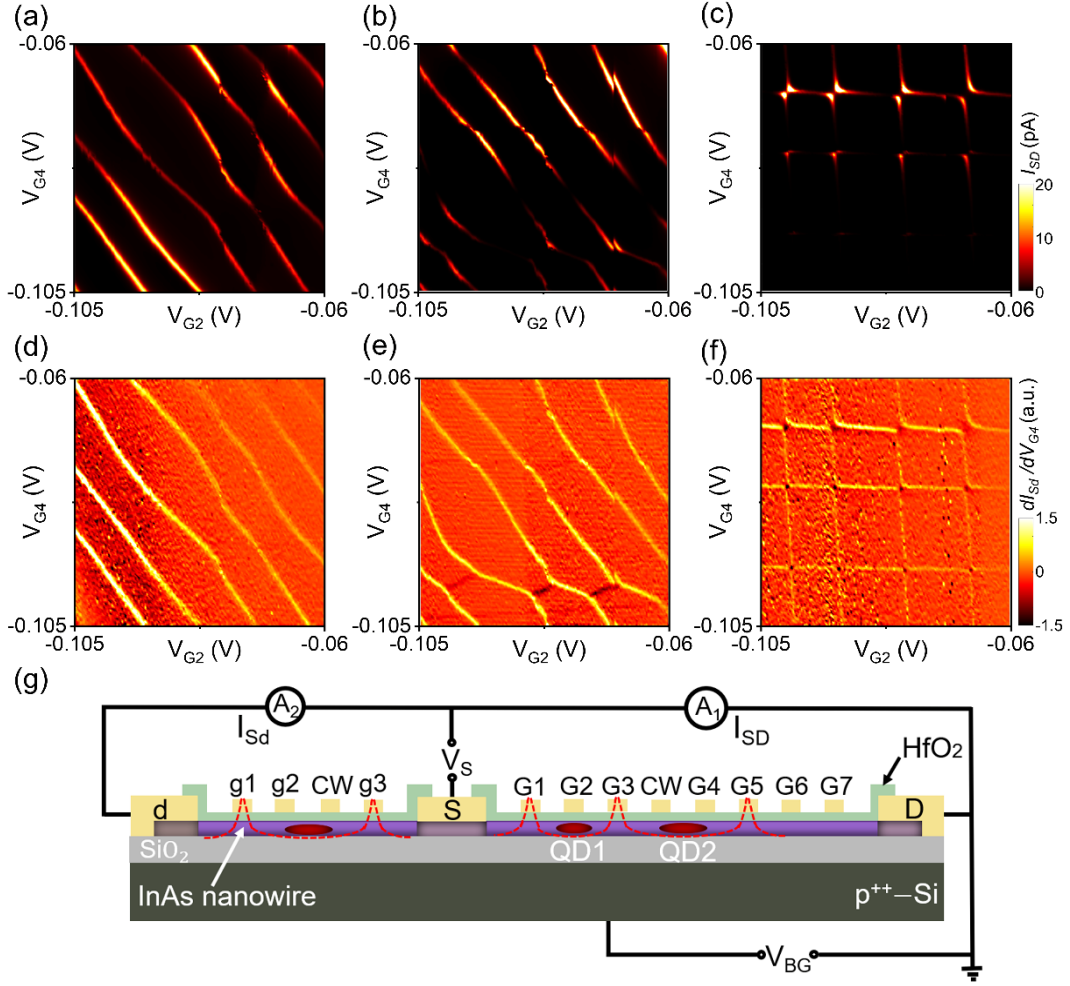

**Figure S3.** (a)-(c) Measured source-drain current $I_{SD}$ of the DQD as defined in (g) at $V_S$ = 70 μV as a function of $V_{G2}$ and $V_{G4}$ (charge stability diagrams) at three different $V_{G3}$. (d)-(f) Corresponding charge stability diagrams of the DQD as detected by the transconductance $dI_{Sd}/dV_{G4}$ measurements of the QD charge sensor. Here, the QD sensor is defined as in (g) by setting $V_{g2}$= 0.108V [i.e., at the declining slope of the same Coulomb current peak as in Figure 3(b) of the main article]. In (a) and (d), straight diagonal charge state transition lines are observed, indicating that a large single QD is formed between gates G1 and G5 at $V_{G3}$ = 0 V. In (b) and (e), bent charge state transition lines are observed, indicating that a DQD with a strong inter-dot coupling is formed at $V_{G3}$ = −0.65 V. In (d) and (f), a DQD with a weak inter-dot coupling is formed by setting $V_{G3}$ = −0.88 V, as regular hexagon-shaped patterns of the current lines are observed. (g) Cross-sectional schematic view of the DQD device and the integrated QD sensor. The DQD was defined by using gates G1 and G5 as the outer tunneling barrier gates, G3 as the inter-dot tunneling barrier gate, and G2 and G4 as the two plunger gates. The locations of the sensor QD and the two QDs in the DQD are marked with red ovals.

The main features seen in the direct transport current measurements shown in Figures S3(a) to S3(c) are accurately reproduced by the transconductance measurements of the QD sensor shown in Figures S3(d) to S3(f). Note that in Figures

S3(d) to S3(f), the charge state transition lines involving transferring charges from the reservoirs to the DQD appear to be bright color lines, while the charge-conserved charge state transition lines, in which only the charge exchange between the two QDs occurs, appear to be dark color lines. Note also that the hardly visible direct transport current lines in Figure S3(c), due to the presence of opaque tunneling barriers in the DQD, are clearly observed in the transconductance measurements of the QD sensor shown in Figure S3(f), verifying an excellent detection sensitivity of the QD sensor to the charge state transitions in the DQD.

**III. Capacitance network model for the TQD studied in the main article**

The charge stability diagram of the TQD studied in the main article is simulated based on an electrostatic capacitance network model. In the model, the TQD is described as a network of tunnel resistors and capacitors. The equivalent circuit diagram, as depicted in Figure 3(c) of the main article, includes nodes of three QDs (QD1, QD2 and QD3), three gates (G2, G3 and G4) that are adjusted in the measurements, and source and drain electrodes. In the simulation, we consider three kinds of major mutual capacitances: (1) the capacitances ($C_{Mij}$) between QD$i$ and QD$j$ ($i, j = 1, 2, 3$ and $j \neq i$), (2) the capacitances ($C_{G3\text{-}QDi}$ and $C_{G4\text{-}QDi}$) between QD$i$ ($i = 1, 2, 3$) and gates G3 and G4 that are swept in the measurements, and (3) the capacitance $C_{G2\text{-}QD1}$ between QD1 and gate G2 that is adjusted to tune the TQD to different regimes of interest. The total charge on each node is denoted as $Q_{QDi}$ ($i = 1, 2, 3$) and the electrostatic potential is denoted as $V_i$ ($i =$ QD1, QD2, QD3, G2, G3, G4, S or D). According to the classical theory, the total charge on each dot can be determined by the relevant gate voltages and capacitances as:

$$\begin{aligned}
Q_{QD1} &= C_S(V_{QD1} - V_S) + C_{G2-QD1}(V_{QD1} - V_{G2}) + C_{G3-QD1}(V_{QD1} - V_{G3}) \\
&\quad + C_{G4-QD1}(V_{QD1} - V_{G4}) + C_{M12}(V_{QD1} - V_{QD2}) + C_{M13}(V_{QD1} - V_{QD3}), \\
Q_{QD2} &= C_{M12}(V_{QD2} - V_{QD1}) + C_{G3-QD2}(V_{QD2} - V_{G3}) + C_{G4-QD2}(V_{QD2} - V_{G4}) \\
&\quad + C_{M23}(V_{QD2} - V_{QD3}), \\
Q_{QD3} &= C_{M23}(V_{QD3} - V_{QD2}) + C_{G4-QD3}(V_{QD3} - V_{G4}) + C_{M13}(V_{QD3} - V_{QD1}) \\
&\quad + C_{G3-QD3}(V_{QD3} - V_{G3}) + C_D(V_{QD3} - V_D).
\end{aligned} \quad (1)$$

Defining the vectors,
$\hat{V}_{QD} = \{V_{QD1}, V_{QD2}, V_{QD3}\}^T, \hat{V}_G = \{V_S, V_{G2}, V_{G3}, V_{G4}, V_D\}^T, \hat{Q}_{QD} = \{Q_{QD1}, Q_{QD2}, Q_{QD3}\}^T$,
Eq. (1) can be written as,

$$\hat{Q}_{QD} = \boldsymbol{C}_{QD} \cdot \hat{V}_{QD} + \boldsymbol{C}_{QD-G} \cdot \hat{V}_G, \quad (2)$$

with matrices $\boldsymbol{C}_{QD}$ and $\boldsymbol{C}_{QD-G}$ given by

$$\boldsymbol{C}_{QD} = \begin{pmatrix} C_1 & -C_{M12} & -C_{M13} \\ -C_{M12} & C_2 & -C_{M23} \\ -C_{M13} & -C_{M23} & C_3 \end{pmatrix}, \quad (3)$$

$$\mathbf{C}_{QD-G} = \begin{pmatrix} -C_S & -C_{G2-QD1} & -C_{G3-QD1} & -C_{G4-QD1} & 0 \\ 0 & 0 & -C_{G3-QD2} & -C_{G4-QD2} & 0 \\ 0 & 0 & -C_{G3-QD3} & -C_{G4-QD3} & -C_D \end{pmatrix}, \qquad (4)$$

where

$$C_1 = C_S + C_{G2-QD1} + C_{G3-QD1} + C_{G4-QD1} + C_{M12} + C_{M13},$$
$$C_2 = C_{M12} + C_{G3-QD2} + C_{G4-QD2} + C_{M23},$$
$$C_3 = C_{M23} + C_{G3-QD3} + C_{G4-QD3} + C_{M13} + C_D. \qquad (5)$$

Then, the electric potentials of the QDs can be calculated as

$$\hat{V}_{QD} = \mathbf{C}_{QD}^{-1}(\hat{Q}_{QD} - \mathbf{C}_{QD-G} \cdot \hat{V}_G), \qquad (6)$$

and the electrostatic energy of the QDs can be obtained from

$$E = \frac{1}{2}\hat{V}_{QD}^T \cdot \mathbf{C}_{QD} \cdot \hat{V}_{QD} = \frac{1}{2}\hat{V}_{QD}^T(\hat{Q}_{QD} - \mathbf{C}_{QD-G} \cdot \hat{V}_G)$$
$$= \frac{1}{2}(\hat{Q}_{QD} - \mathbf{C}_{QD-G} \cdot \hat{V}_G)^T [\mathbf{C}_{QD}^{-1}]^T (\hat{Q}_{QD} - \mathbf{C}_{QD-G} \cdot \hat{V}_G). \qquad (7)$$

As $Q_{QDi} = -eN_i$ ($i = 1, 2, 3$), the chemical potentials $\mu_i$ in individual QDs are

$$\mu_1(N_1, N_2, N_3, V_{G3}, V_{G4}, V_{G5}) = E(N_1, N_2, N_3, V_{G3}, V_{G4}, V_{G5}) -$$
$$E(N_1 - 1, N_2, N_3, V_{G3}, V_{G4}, V_{G5}),$$
$$\mu_2(N_1, N_2, N_3, V_{G3}, V_{G4}, V_{G5}) = E(N_1, N_2, N_3, V_{G3}, V_{G4}, V_{G5}) -$$
$$E(N_1, N_2 - 1, N_3, V_{G3}, V_{G4}, V_{G5}),$$
$$\mu_3(N_1, N_2, N_3, V_{G3}, V_{G4}, V_{G5}) = E(N_1, N_2, N_3, V_{G3}, V_{G4}, V_{G5}) -$$
$$E(N_1, N_2, N_3 - 1, V_{G3}, V_{G4}, V_{G5}). \qquad (8)$$

The electron occupancies of the three QDs can be manipulated by adjusting the voltages of the barrier gates and/or plunger gates. When the occupancy number of QD$i$ increases from $N_i$ to $N_i + 1$, the change in the voltage on gate G$j$ can be derived based on the equation of

$$\mu_i(N_1, V_{Gj}) = \mu_i(N_1 + 1, V_{Gj} + \Delta V_{Gj-QDi}). \qquad (9)$$

Thus, with the use of Eqs. (7) and (8), the corresponding changes in the voltages on the swept gates G3 and G4 can be calculated as

$$\Delta V_{G3-QD1} = \frac{e}{C_{G3-QD1} + \frac{C_{M12}}{C_2}C_{G3-QD2} + \frac{C_{M13}}{C_3}C_{G3-QD3}},$$

$$\Delta V_{G4-QD1} = \frac{e}{C_{G4-QD1} + \frac{C_{M12}}{C_2}C_{G4-QD2} + \frac{C_{M13}}{C_3}C_{G4-QD3}},$$

$$\Delta V_{G3-QD2} = \frac{e}{C_{G3-QD2} + \frac{C_{M12}}{C_1}C_{G3-QD1} + \frac{C_{M23}}{C_3}C_{G3-QD3}},$$

$$\Delta V_{G4-QD2} = \frac{e}{C_{G4-QD2} + \frac{C_{M12}}{C_1}C_{G4-QD1} + \frac{C_{M23}}{C_3}C_{G4-QD3}},$$

$$\Delta V_{G3-QD3} = \frac{e}{C_{G3-QD3} + \frac{C_{M13}}{C_1}C_{G3-QD1} + \frac{C_{M23}}{C_2}C_{G3-QD2}},$$

$$\Delta V_{G4-QD3} = \frac{e}{C_{G4-QD3} + \frac{C_{M13}}{C_1} C_{G4-QD1} + \frac{C_{M23}}{C_2} C_{G4-QD2}}. \quad (10)$$

The capacitances $C_S$ and $C_D$ are estimated from the Coulomb diamond measurements of single QDs as shown in Figs. S2(a)-S2(b). Based on the charge stability diagram shown in Figures 3(a) and 3(b) of the main article, the values of $\Delta V_{Gj-QDi}$ can be obtained as $\Delta V_{G3-QD1}$= 0.0228 V, $\Delta V_{G4-QD1}$= 0.075 V, $\Delta V_{G3-QD2}$ = 0.0245 V, $\Delta V_{G4-QD2}$= 0.0277 V, $\Delta V_{G3-QD3}$= 0.143 V, and $\Delta V_{G4-QD3}$= 0.0285 V. Substituting these values of voltage changes into Eq. (10) and considering the fact that $C_{i=1,2,3} \gg C_{Mij}$ ($ij$=12, 13, 23), the values of all the remaining capacitances can be extracted. All the capacitance parameters used to obtain the simulated charge stability diagram of the TQD, as shown in Figure 3(d) of the main article, are listed in Table S1.

**Table S1.** Values of the capacitances, in units of aF, used to simulate the charge stability diagram of the TQD shown in Figure 3(b) of the main article. The simulated charge stability diagram for the TQD device is shown in Figure 3(d) of the main article.

| $C_S$ | $C_D$ | $C_{G3-QD1}$ | $C_{G4-QD1}$ | $C_{M12}$ | $C_{M13}$ |
|---|---|---|---|---|---|
| 9.80 | 8.00 | 6.98 | 2.12 | 1.28 | 0.32 |
| $C_{G3-QD2}$ | $C_{G4-QD2}$ | $C_{M23}$ | $C_{G3-QD3}$ | $C_{G4-QD3}$ | $C_{G2-QD1}$ |
| 5.80 | 3.76 | 1.60 | 0.63 | 5.28 | 4.80 |